\documentclass[12pt,showpacs,showkeys,amsmath,amssymb]{revtex4}

\usepackage{graphicx}
\usepackage{dcolumn}
\usepackage{bm}

\newcommand{\diag}{\operatorname{diag}}

\begin{document}

\title{Nonlocal Special Relativity}
\author{Bahram Mashhoon}
\affiliation{Department of Physics and Astronomy\\University of Missouri-Columbia\\Columbia, Missouri 65211, USA}
\email{mashhoonb@missouri.edu}

\begin{abstract}
In the special theory of relativity, Lorentz invariance is extended in Minkowski spacetime from ideal inertial observers to actual observers by means of the hypothesis of locality, which postulates that accelerated observers are always pointwise inertial. A critical examination of the locality assumption reveals its domain of validity: it is true for pointwise coincidences, but is in conflict with wave-particle duality. To remedy this situation, a nonlocal theory of accelerated systems is presented that reduces to the standard theory in the limit of small accelerations. Some of the main consequences of nonlocal special relativity are briefly outlined. 
\end{abstract}

\keywords{Relativity, nonlocality, accelerated observers.}
\pacs{03.30.+p, 11.10.Lm, 04.20.Cv}

\maketitle

\section{Introduction}\label{s1}

Minkowski's idea of putting time and space on an equal footing thereby effecting their conceptual unification in spacetime was a fundamental step in the development of the special theory of relativity~\cite{27}. Moreover, Minkowski's emphasis on spacetime geometry paved the way for the subsequent emergence of Einstein's geometric theory of the gravitational field~\cite{28}. 

The special theory of relativity---that is, the standard relativistic physics of Minkowski spacetime---is primarily based on a fundamental symmetry in nature, namely, \emph{Lorentz invariance}. Lorentz invariance relates the physical measurements of inertial observers at rest in one inertial frame to those at rest in another inertial frame. The basic laws of microphysics have been formulated with respect to inertial observers; however, all actual observers are accelerated. The term ``observer'' is here used in an extended sense to include any measuring device. To interpret experimental results, it is therefore necessary to extend Lorentz invariance to accelerated observers. That is, a physical connection must be established between accelerated and inertial observers. The assumption that is employed in the standard theory of relativity is the \emph{hypothesis of locality}, which asserts that an accelerated observer is pointwise physically equivalent to an otherwise identical momentarily comoving inertial observer~\cite{29}-\cite{31}. Thus, Lorentz invariance and the hypothesis of locality together form the physical basis for the special theory of relativity. Lorentz invariance is consistent with quantum theory, but, as will become clear in section~\ref{s2}, this is not the case with the hypothesis of locality. The aim of the nonlocal formulation of special relativity is to correct this situation. 

In Minkowski's treatment of special relativity, the locality assumption was highlighted as a fundamental axiom (see page 80 of~\cite{27}):

\begin{quote}\emph{``The substance at any world-point may always, with the appropriate determination of space and time, be looked upon as at rest.''}\end{quote}

The physical origin and limitations of the hypothesis of locality are critically examined in the first part of this paper. The second part starts with a generalization of the hypothesis of locality that is consistent with wave-particle duality. On this basis, a nonlocal theory of accelerated systems is presented that goes beyond the hypothesis of locality, but reduces to the standard theory in the limit of small accelerations. The main features of this nonlocal special relativity are briefly examined. 

\section{Hypothesis of locality}\label{s2}

 In the special theory of relativity, Lorentz invariance is extended to
accelerated systems by means of the following assumption.

\vspace{14pt} POSTULATE OF LOCALITY: An \emph{accelerated} observer (measuring device)
along its worldline is at each instant physically equivalent to a
hypothetical \emph{inertial} observer (measuring device) that is otherwise
identical and instantaneously comoving with the accelerated observer
(measuring device).\vspace{14pt}

\noindent What is the physical basis of this supposition? The locality assumption originates from Newtonian mechanics of point particles: the accelerated observer and the otherwise identical instantaneously comoving inertial observer have the same position and velocity; hence, they share the same state and are thus pointwise physically identical in classical mechanics. Therefore, in the treatment of accelerated systems of reference in Newtonian mechanics, no new physical assumption is required. The hypothesis of locality should hold equally well in the classical relativistic mechanics of point particles. That is, if all physical phenomena could be reduced to \emph{pointlike coincidences} of classical particles and rays of radiation, then the hypothesis of locality would be exactly valid.

In the special theory of relativity, an accelerated observer is in effect replaced---on the basis of the hypothesis of locality---by a continuous infinity of hypothetical momentarily comoving inertial observers. It seems that Lorentz first introduced such an assumption in his theory of electrons in order to ensure that an electron, conceived as a small ball of charge, would always be Lorentz contracted along its direction of motion~\cite{32}. It was clearly recognized by Lorentz that this is simply an approximation whose validity rests on the supposition that the electron velocity would change over a time scale that is much longer than the period of internal oscillations of the electron (see section 183 on page 216 of~\cite{32}). 

A similar assumption was simply adopted by Einstein for rods and clocks (see the footnote on page 60 of~\cite{28}). Indeed, the hypothesis of locality underlies Einstein's development of the theory of relativity. For instance, the locality assumption fits perfectly together with Einstein's local principle of equivalence to ensure that every observer in a gravitational field is pointwise inertial. In fact, to preserve the operational significance of Einstein's heuristic principle of equivalence---namely, the presumed local equivalence of an observer in a gravitational field with an accelerated observer in Minkowski spacetime---it must be coupled with a statement regarding what accelerated observers actually measure. When coupled with the hypothesis of locality, Einstein's principle of equivalence provides a physical basis for a field theory of gravitation that is consistent with (local) Lorentz invariance. 

Following Einstein's development of the general theory of relativity, Weyl discussed the physical basis for the hypothesis of locality (see pages 176-177 of~\cite{33}). In particular, Weyl noted that the locality hypothesis was an adiabaticity assumption analogous to the one for sufficiently slow processes in thermodynamics and would therefore be expected to be a good approximation only up to some acceleration~\cite{33}.

The modern experimental basis of the theory of relativity necessitates a discussion of the locality assumption within the context of the quantum theory of measurement. The standard quantum theory of measurement deals with \emph{ideal} inertial measuring devices. An \emph{actual} measuring device is in general noninertial and has specific limitations---due to the nature of its construction and its modes of operation---that must obviously be taken into account in any actual experiment. The fact that these experimental limitations exist clearly does not invalidate the standard quantum measurement theory. This circumstance illustrates the general relationship between the basic theoretical structure of a physical theory and the additional limitations of actual laboratory devices. Therefore, throughout this work, we assume that the measuring devices are ideal. The response of the ideal measuring devices to acceleration---that is, the influence of inertial effects on their operation---should eventually be determined on the basis of a proper theory of accelerated systems. 

A physical theory of space and time must deal with the issue of whether ideal rods and clocks satisfy the locality assumption. Measuring devices that do so are usually called ``standard''. Thus the clock hypothesis states that standard clocks measure proper time. The inertial effects in a standard device must therefore operate in just such a way as to make a standard device always locally inertial. This is clearly a physical assumption regarding ideal accelerated devices. In this connection, it is interesting to mention here a remark by Sommerfeld in his notes on Minkowski's 1908 paper (see page 94 of~\cite{27}):

\begin{quote}``...The assertion is based, as Einstein has pointed out, on the unprovable assumption that the clock in motion actually indicates its own proper time, i.e. that it always gives the time corresponding to the state of velocity, regarded as constant, at any instant. The moving clock must naturally have been moved with acceleration (with changes of speed or direction) in order to be compared with the stationary clock at the world-point $P$....''\end{quote}

An accelerated observer is characterized by the translational acceleration of its worldline and the rotation of its spatial frame. Using these quantities and the speed of light in vacuum, one can construct an acceleration length $\ell$ that is characteristic of the scale of spatial variation of the state of the observer. If $\lambda$ is the intrinsic length that is characteristic of the phenomenon under observation, then we expect deviations from the hypothesis of locality of order $\lambda /\ell$. In an Earth-based laboratory, for instance, the main translational and rotational acceleration lengths would be $c^2/g_\oplus \approx 1$ light year and $c/\Omega _\oplus \approx 28$ A.U., respectively. Thus in most experimental situations $\lambda /\ell$ is negligibly small and the hypothesis of locality is a good approximation. 

To illustrate the consistency of these ideas, consider, for instance, a classical charged particle of mass $m$ and charge $\epsilon$ that is accelerated by an external force $\mathbf{f}$. The characteristic wavelength of the electromagnetic waves radiated by the accelerated charge is $\lambda \sim \ell$. Thus $\lambda /\ell \sim 1$ in the interaction of the particle with the electromagnetic field and a significant breakdown of the locality assumption is expected in this case. That is, the state of the radiating charged particle cannot be specified by its position and velocity alone. This conclusion is indeed consistent with the Abraham-Lorentz equation
\begin{equation}\label{eq:114} m\frac{d\mathbf{v}}{dt}-\frac{2}{3}\frac{\epsilon^2}{c^3} \frac{d^2\mathbf{v}}{dt^2}+\cdots =\mathbf{f},\end{equation}
which describes the motion of the particle in the nonrelativistic approximation. 

A second example involves the dilation of muon lifetime when muon decay is measured in a storage ring~\cite{34,35}. The hypothesis of locality implies that $\tau_\mu =\gamma\tau^0_\mu$, where  $\tau^0_\mu$ is the lifetime of the muon at rest and $\gamma$ is the Lorentz factor corresponding to the circular motion of muons in the storage ring. To avoid the locality assumption, one can suppose that the muon decays from a high-energy Landau level in a constant magnetic field. This approach has been followed in~\cite{36} for the calculation of muon decay on the basis of the quantum theory. The result can be expressed as~\cite{37}
\begin{equation}\label{eq:115} \tau_\mu \approx \gamma \tau^0_\mu \left[ 1+\frac{2}{3}\left( \frac{\lambda_C}{\ell}\right)^2\right],\end{equation}
where $\lambda_C$ is the Compton wavelength of the muon and $\ell$ is its effective acceleration length. That is, $\ell=c^2/g$, where $g\sim 10^{18}g_\oplus$ is the effective centripetal acceleration of the muon in the storage ring. In practice, the nonlocal correction term turns out to be very small ($\sim 10^{-25}$).

It follows from the foregoing examples that the hypothesis of locality is in general inadequate for wave phenomena. These considerations provide the motivation to develop a theory of accelerated systems that goes beyond the hypothesis of locality and is consistent with wave-particle duality. 

\section{Accelerated observers}\label{s3}

Imagine a global inertial frame $\mathcal{S}$ in Minkowski spacetime with coordinates $x^\alpha =(ct,\mathbf{x})$. Throughout this paper, we assume that these inertial coordinates are employed by \emph{all} observers in $\mathcal{S}$.  We choose units such that $c=\hbar=1$; moreover, in our convention, the signature of the Minkowski metric is $+2$, except in sections~\ref{sec:8} and \ref{sec:9} as well as Appendix~\ref{sec:A}, where it is $-2$. An accelerated observer with proper time $\tau$ and worldline $ x^\alpha (\tau )$ passes through a continuum of instantaneously comoving inertial observers. Each inertial observer in $\mathcal{S}$ is endowed with natural temporal and spatial axes that constitute its orthonormal tetrad. It then follows from the locality assumption that an accelerated observer carries an orthonormal tetrad $\lambda^\mu_{\;\;(\alpha )}(\tau )$ along its worldline such that at each instant of proper time $\tau$ the observer's tetrad coincides with that of the momentarily comoving inertial observer. We note that its $4$-velocity $\lambda^\mu_{\;\; (0)}=dx^\mu /d\tau$ is a unit vector tangent to the worldline and acts as the local temporal axis of the observer, while its local spatial frame is defined by the unit spacelike axes $\lambda^\mu_{\;\;(i)}$, $i=1,2,3$. To avoid unphysical situations, it is natural to assume that the observer's acceleration lasts only for a finite interval of time. 

The theoretical role of the hypothesis of locality in the measurement of spatial and temporal intervals by a congruence of accelerated observers has been scrutinized in a number of studies. It has been shown that a consistent treatment is possible only when the distances involved are very small compared to the relevant acceleration lengths~\cite{30,31,38}. In particular, if $D$ is the spatial dimension of a standard measuring device, then $D\ll\ell$. The various implications of this result for quantum measurement theory have been discussed in~\cite{29,37}. Further studies of the function of locality in the theory of relativity are contained in~\cite{39}-\cite{44} and the references therein. In this paper, we assume that the accelerated observer has access to standard measuring devices for the determination of its proper time and its local tetrad frame. 

The accelerated observer's tetrad frame varies along its worldline in accordance with
\begin{equation}\label{eq:116} \frac{d\lambda^\mu_{\;\;(\alpha)}}{d\tau }=\phi _\alpha^{\;\;\beta}\lambda^\mu_{\;\;(\beta)}.\end{equation}
Here $\phi_{\alpha\beta} =-\phi_{\beta \alpha}$ is the invariant antisymmetric acceleration tensor of the observer. This tensor can be decomposed as $\phi_{\alpha \beta}\mapsto (-\mathbf{g},\boldsymbol{\Omega})$ in close analogy with the Faraday tensor. The translational acceleration of the observer is represented by the ``electric'' part of the tensor ( $\phi_{0i}=g_i$), while the angular velocity of the rotation of the observer's spatial frame with respect to a locally nonrotating (i.e. Fermi-Walker transported) frame is given by the ``magnetic'' part of the tensor ($\phi_{ij}=\epsilon_{ijk}\Omega ^k$) . 

An accelerated observer is in general characterized by its worldline as well as the associated tetrad frame. The local rate of variation of the state of the accelerated observer is thus completely specified by the six coordinate scalars $\mathbf{g}$ and $\boldsymbol{\Omega}$. These components of the acceleration tensor and their temporal derivatives in general enter into the definition of the relevant acceleration lengths of the observer~\cite{30}.

\section{Beyond locality}\label{s4}

Consider the reception of electromagnetic waves by an accelerated observer in the global inertial frame$\;\mathcal{S}$. To measure the frequency of the incident wave, the observer needs to register several oscillations before an adequate determination of the frequency becomes even possible. Thus an extended period of proper time is necessary for this purpose. On the other hand, for an incident wave with propagation vector $k^\alpha=(\omega ,\mathbf{k})$, the hypothesis of locality implies that at each instant of proper time $\tau$, the observer measures $\omega_D(\tau )=-k_\alpha \lambda^\alpha_{\;\;(0)}$ via the Doppler effect. Thus from a physical standpoint, such an instantaneous Doppler formula can be strictly valid only in the eikonal limit of rays of radiation. 

There is an alternative way to apply the locality hypothesis to this situation~\cite{45}-\cite{48}. One can employ the instantaneous Lorentz transformations from $\mathcal{S}$ to the local inertial frame of the observer at $\tau$ to determine the electromagnetic field that is presumably measured by the accelerated observer at $\tau$. The Fourier analysis of this measured field in terms of proper time would then result in the invariant frequency $\hat{\omega}$ measured by the accelerated observer. We note that the measured field in this case can be directly evaluated by the projection of the Faraday tensor of the wave in $\mathcal{S}$ on the tetrad frame of the accelerated observer. This approach is nonlocal insofar as it relies on Fourier analysis, but the field determination is still based on the locality assumption. However, as pointed out by Bohr and Rosenfeld~\cite{49,50}, it is not physically possible to measure an electromagnetic field at one event; instead, a certain averaging process is required. To remedy this situation, nonlocal field determination is considered in the next section in connection with the fully nonlocal theory of accelerated observers. 

In the rest of this section, it proves instructive to explore briefly the physical difference between $\hat{\omega}$ and $\omega_D$ for the specific case of a uniformly rotating observer. Imagine an observer rotating in the positive sense with constant frequency $\Omega_0$ about the direction of propagation of a plane monochromatic electromagnetic wave of frequency $\omega$. The observer follows a circle of radius $\rho$ with uniform speed $v=\rho\Omega_0$ in a plane and the wave is normally incident on this plane. In this case, $\omega_D=\gamma\omega$ by the transverse Doppler effect. Here the Lorentz factor takes due account of time dilation. On the other hand, the Fourier analysis of the field, pointwise ``measured'' by the observer via the hypothesis of locality, reveals that
\begin{equation}\label{eq:117}\hat{\omega}=\gamma (\omega \mp \Omega_0).\end{equation}
Here, the upper (lower) sign refers to incident positive (negative) helicity radiation. The new result reflects time dilation as well as the coupling of photon helicity with the rotation of the observer. The nonlocal aspect of the helicity-rotation coupling is evident from $\hat{\omega}=\omega_D(1\mp \Omega_0/\omega)$, where $\Omega_0/\omega$ is the ratio of the reduced wavelength of the wave ($1/\omega$) to the acceleration length of the observer ($1/\Omega_0$). It is possible to give a simple intuitive explanation for the helicity-rotation coupling. In a positive (negative) helicity wave, the electric and magnetic field vectors rotate with frequency $\omega\; (-\omega)$ about the direction of wave propagation; however, from the viewpoint of the observer, the electric and magnetic fields rotate with relative frequency $\omega-\Omega_0\; (-\omega-\Omega_0$) if time dilation is ignored. Taking time dilation into account, one recovers the formula for $\hat{\omega}$. 

There exists ample observational evidence for the spin-rotation coupling~\cite{51}-\cite{54}. In particular, it accounts for the phenomenon of \emph{phase wrap-up} in the GPS. Indeed, for $\gamma \ll 1$ and $\Omega_0\ll\omega$, the frequency $\hat{\omega} \approx \omega \mp\Omega_0$ has been verified with $\omega /(2\pi)\sim 1$ GHz and $\Omega_0/(2\pi)\sim 8$ Hz~\cite{55}. It would be interesting to investigate the validity of equation \eqref{eq:117} beyond the eikonal regime ($\omega \gg \Omega_0$) that is experimentally accessible at present. 

For incident radiation of positive helicity, the frequency measured by the rotating observer could be negative for $\omega <\Omega_0$ or zero for $\omega =\Omega_0$. The former does not pose a basic difficulty once it is recognized that the notion of relativity does not extend to accelerated observers~\cite{45}. The situation is different, however, with the latter possibility. By a mere rotation, observers could in principle stay at rest with an electromagnetic wave. That is, the electromagnetic field is oscillatory in space but has no temporal dependence with respect to all observers that rotate uniformly with frequency $\omega$ about the direction of incidence~\cite{56}. An analogous problem plagued the pre-relativistic Doppler formula and this circumstance influenced Einstein's path to relativity theory (see page 53 of Einstein's autobiographical notes in~\cite{57}). For accelerated observers, this fundamental difficulty is associated with the pointwise determination of the electromagnetic field and its avoidance plays a significant part in nonlocal special relativity. 

For the general case of oblique incidence of a basic radiation field on the plane of the observer's circular orbit, the expression for the spin-rotation coupling turns out to be $\hat{\omega}=\gamma (\omega -M\Omega_0)$, where $M=0$, $\pm1,\pm 2,\dots$ for a scalar or a vector field, while $M\mp\frac{1}{2}=0$, $\pm 1,\pm 2,\dots$ for a Dirac field. Here $M$ is the component of the total angular momentum of the radiation field along the axis of rotation of the observer. The measured frequency can be negative for $\omega <M\Omega_0$ and vanishes for $\omega =M\Omega_0$. In the eikonal approximation ($\omega \gg\Omega_0$), one can show that 
\begin{equation}\label{eq:118} \hat{\omega}\approx \gamma (\omega -\mathbf{v}\cdot\mathbf{k})-\gamma\; \mathbf{s}\cdot \boldsymbol{\Omega}_0,\end{equation}
where  $\mathbf{s}$ is the intrinsic spin of the particle. The general phenomenon of spin-rotation coupling is a manifestation of the inertia of intrinsic spin~\cite{58}-\cite{71}. A more complete discussion as well as list of references is contained in~\cite{72}-\cite{75}. We now turn to the issue of nonlocal field determination by accelerated observers.

\section{Acceleration-induced nonlocality}\label{s5}

Consider a basic radiation field $\psi (x)$ in a background global inertial frame $\mathcal{S}$ in Minkowski
spacetime and an accelerated observer that measures this field.  The events along the worldline of the observer are
characterized by its proper time $\tau$.  The observer passes through an infinite sequence of hypothetical momentarily
comoving inertial observers.  Let  $\hat{\psi}(\tau )$ be the field measured by these inertial observers.  The relation
between $\hat{\psi}$  and $\psi$  can be determined from the fact that the local inertial spacetime of the comoving
inertial observer at $\tau$---i.e. the global inertial frame in which it is at rest---is related to $\mathcal{S}$ by a
Poincar\'e transformation $x'=Lx+a$.  It follows that  $\psi '(x')=\Lambda (L)\psi (x)$, where $\Lambda$ belongs to a
matrix representation of the Lorentz group.  For instance, $\Lambda$ is unity for a scalar field. 
Thus  $\hat{\psi}(\tau )=\Lambda (\tau )\psi (\tau)$ along the worldline of the accelerated observer. 

The fundamental laws of microphysics have been formulated with respect to inertial observers.  On the other hand,
physical measurements are performed by observers that are, in general, accelerated.  To interpret observation via
theory, a connection must be established between the field $\hat{\Psi}(\tau)$ that is actually measured by the
accelerated observer and $\hat{\psi}(\tau)$.
 The standard theory of relativity based on the hypothesis of locality postulates that the accelerated observer is
pointwise inertial and hence $\hat{\Psi}(\tau)=\hat{\psi}(\tau)$.  This is, of course, the simplest
possibility and has been quite successful as a first approximation.  It is consistent with the physical principles of
superposition and causality.  It is important, however, to go beyond this relation in view of the limitations of the
hypothesis of locality.  The most general linear relationship between $\hat{\Psi}(\tau)$ and $\hat{\psi}(\tau)$ that preserves
linearity and causality is
\begin{equation}\label{eq:1} \hat{\Psi}(\tau)=\hat{\psi}(\tau)+\int^{\tau}_{\tau_{0}}\hat{K}(\tau ,\tau')\hat{\psi}(\tau
')d\tau ',\end{equation}
where $\tau_{0}$ is the instant at which the observer's acceleration is turned on and the kernel $\hat{K}$ vanishes in the
absence of acceleration.  The ansatz \eqref{eq:1} goes beyond the locality assumption by virtue of an integral over the
past
worldline of the observer.  Qualitatively, the contribution of the nonlocal part of \eqref{eq:1} can be estimated to be vanishingly small for $\lambda /\ell\to 0$, where $\lambda$ is a typical wavelength and $\ell$ is the acceleration length, in general agreement with expected deviations from the hypothesis of locality.  Thus \eqref{eq:1} has been adopted as the basic assumption of the nonlocal theory of accelerated systems \cite{1,2}.  Nonlinear generalizations of \eqref{eq:1} may be contemplated, of course, but these appear unnecessary at the present stage of development.  Equation \eqref{eq:1} has the form of a Volterra integral equation of the second kind.  Therefore, the relationship between $\hat{\Psi}$ and $\hat{\psi}$ is unique in the space of continuous functions in accordance with Volterra's theorem~\cite{3}.  This uniqueness result---which plays an important role in the determination of the kernel $\hat{K}$---has been extended to the Hilbert space of square-integrable functions by Tricomi~\cite{4}.  For the physical fields under consideration here, we assume that $\hat{\psi}$ is indeed uniquely determined by $\hat{\Psi}$; that is, 
\begin{equation}\label{eq:2} \hat{\psi} (\tau )=\hat{\Psi}(\tau )+\int^\tau_{\tau_0} \hat{R} (\tau ,\tau ')\hat{\Psi} (\tau ')d\tau ',\end{equation}
where $\hat{R}$ is the resolvent kernel \cite{3}-\cite{5}.

Equation~\eqref{eq:1} is reminiscent of the nonlocal characterization of certain constitutive properties of continuous media that exhibit memory-dependent phenomena (``after-effects'').   This subject has a long history (see, for instance, \cite{5}).  It is important to remark that the nonlocality considered in the present work is in the absence of any medium; rather, it is associated with the vacuum state as perceived by accelerated observers \cite{6}.

How should the kernel $\hat{K}$ be determined?  This turns out to be a rather complicated issue as discussed in the next section; however, a key aspect of the nonlocal theory should be the resolution of the basic difficulty encountered in the discussion of the spin-rotation coupling.  We, therefore, raise a consequence of Lorentz invariance, that an inertial observer cannot stay at rest with respect to a fundamental radiation field, to the level of a postulate that must hold for all observers.  Thus a basic radiation field cannot stand completely still with respect to an accelerated observer \cite{1,2}.  To implement this requirement in the nonlocal theory, let us first recall an aspect of the Doppler formula in electrodynamics for an inertial observer moving with uniform velocity $\mathbf{v}\!:\omega '=\gamma (\omega -\mathbf{v}\cdot \mathbf{k})$, where $\omega =|\mathbf{k}|$.  We note that $\omega '=0$ only when $\omega =0$, since $v<1$; that is, if the moving observer encounters a constant wave field, then the field must already be constant for observers at rest.  The generalization of this circumstance to accelerated observers would imply that if $\hat{\Psi}$ in \eqref{eq:1} turns out to be constant, then $\psi$ must have been a constant field in the first place.  It would then follow from the Volterra-Tricomi uniqueness theorem that for a realistic variable field $\psi (x)$, the measured field $\hat{\Psi}(\tau )$ would never be a constant.  In this way, a basic radiation field can never stand completely still with respect to any observer \cite{1,2}.

Our physical postulate leads to an integral equation for the kernel $\hat{K}$ by means of the nonlocal ansatz \eqref{eq:1}.  Writing $\hat{\psi }(\tau )=\Lambda (\tau )\psi (\tau )$ in \eqref{eq:1} and noting that $\hat{\Psi}(\tau _0)=\hat{\psi}(\tau_0)$, we find that 
\begin{equation}\label{eq:3} \Lambda (\tau _0)=\Lambda (\tau )+\int^\tau_{\tau_0} \hat{K}(\tau ,\tau')\Lambda (\tau ')d\tau ',\end{equation}
once we let $\psi$ and $\hat{\Psi}$ be constants.  Given $\Lambda (\tau )$, this equation is not sufficient to determine the kernel uniquely.  To proceed, a simplifying assumption would be appropriate.  Two possibilities appear natural: (i) $\hat{K}(\tau ,\tau ')$ is only a function of $\tau-\tau'$ or (ii)  $\hat{K}(\tau ,\tau ')$ is only a function of $\tau'$.  These lead to the same constant kernel for uniform acceleration.  The convolution kernel in case (i) was initially adopted in analogy with nonlocal theories of continuous media \cite{1,2}, but was later found to lead to divergences in cases of nonuniform acceleration \cite{7}.  A detailed investigation \cite{8,9} reveals that case (ii) provides the only physically acceptable solution of \eqref{eq:3}, so that 
\begin{equation}\label{eq:4} \hat{K}(\tau ,\tau ')=\hat{k}(\tau ').\end{equation}
Differentiation of \eqref{eq:3} in this case results in 
\begin{equation}\label{eq:5} \hat{k}(\tau )=-\frac{d\Lambda (\tau )}{d\tau} \Lambda^{-1} (\tau ).\end{equation}
This kernel is directly proportional to the acceleration of the observer.  Hence, it follows that once the acceleration is turned off at  $\tau_f $, then for  $\tau >\tau_f$, though the motion of the observer is uniform, there is a constant nonlocal contribution to the measured field in \eqref{eq:1} that is simply a memory of past acceleration.  This constant memory is measurable in principle, but it is simply canceled in a measuring device whenever the device is reset.

With the kernel as in \eqref{eq:4}, the nonlocal part of our main ansatz \eqref{eq:1} takes the form of a weighted average over the past worldline of the accelerated observer such that the  weighting function is directly proportional to the acceleration.  This circumstance is consistent with the Bohr-Rosenfeld viewpoint regarding field determination.  Moreover, this general approach to acceleration-induced nonlocality appears to be consistent with quantum theory \cite{10}.

Substitution of \eqref{eq:4} and \eqref{eq:5} in \eqref{eq:1} results in 
\begin{equation}\label{eq:6}\hat{\Psi} (\tau )=\hat{\psi}(\tau_0) -\int^\tau_{\tau_0} \Lambda (\tau ')\frac{d\psi (\tau ')}{d\tau '} d\tau '.\end{equation}
It follows from this relation that if the field $\psi (x)$ evaluated along the worldline of the accelerated observer turns out to be a constant over a certain interval $(\tau _0,\tau_1)$, then the variable nonlocal part of \eqref{eq:6} vanishes in this interval and the measured field  turns out to be a constant as well for $\tau_0<\tau <\tau_1$.  This result plays an important role in the development of nonlocal field theory of electrodynamics in the next section.  That is, we have thus far considered radiation fields; however, in the next section we need to deal with limiting situations such as electrostatics and magnetostatics as well.

The discussion of spin-rotation coupling in the previous section leads to the conclusion that for a basic scalar field of frequency $\omega$, an observer rotating uniformly with frequency $\Omega_0$ measures $\hat{\omega }=\gamma (\omega -M\Omega_0)$, where $M=0, \pm 1, \pm 2, \dots$. Thus $\hat{\omega }=0$ for $\omega =M\Omega_0$, so that the scalar radiation field is oscillatory in space but stands completely still with respect to the rotating observer.  This possibility is ruled out by the nonlocal theory.  However, for a scalar field $\Lambda=1$ and it follows from (5) that $\hat{k}=0$, so that a basic scalar radiation field is purely local.  Our main postulate therefore implies that a pure scalar (or pseudoscalar) radiation field does not exist.  Nevertheless, scalar or pseudoscalar fields can be composites formed from other basic fields.  There is no trace of a fundamental scalar or pseudoscalar field in the present experimental data; hence, this important implication of nonlocal theory is consistent with observation. 

\section{Nonlocal electrodynamics}\label{sec:6}

Consider the determination of an electromagnetic field by an accelerated observer. According to the fundamental inertial observers in $\mathcal{S}$, there is a vector potential $A_\mu$ and the corresponding Faraday tensor $F_{\mu\nu}$,
\begin{equation}\label{eq:7} F_{\mu\nu}=\partial_\mu A_\nu -\partial_\nu A_\mu.\end{equation}
The associated sequence of locally comoving inertial observers along the worldline of the accelerated observer measure
\begin{equation} \label{eq:8}\hat{A}_\alpha =A_\mu \lambda^\mu_{\;\;(\alpha )},\quad \hat{F}_{\alpha \beta}=F_{\mu\nu}\lambda^\mu_{\;\;(\alpha )}\lambda^\nu _{\;\;(\beta)}.\end{equation}
Our basic ansatz \eqref{eq:1} would then imply that the fields determined by the accelerated observer are
\begin{align}\label{eq:9} \hat{\mathcal{A}}_\alpha (\tau )&=\hat{A}_\alpha (\tau )+\int^\tau _{\tau_0}\hat{K}_\alpha^{\;\;\beta} (\tau ,\tau ')\hat{A}_\beta (\tau ')d\tau ',\\
\label{eq:10} \hat{\mathcal{F}}_{\alpha\beta}(\tau )&= \hat{F}_{\alpha \beta} (\tau )+\int^\tau_{\tau_0} \hat{K}_{\alpha \beta}^{\;\;\;\; \gamma \delta} (\tau ,\tau ')\hat{F}_{\gamma \delta} (\tau ')d\tau '.\end{align}
The gauge dependence of $A_\mu$ translates into the gauge dependence of $\hat{\mathcal{A}}_\alpha$, while $\hat{\mathcal{F}}_{\alpha\beta}$ is gauge invariant.

The simplest choice for the kernels in \eqref{eq:9} and \eqref{eq:10} would be the standard kernels given by \eqref{eq:5}. For the vector potential, this would mean that
\begin{equation}\label{eq:11} \hat{K}_\alpha^{\;\;\beta} (\tau ,\tau ')=-\phi_\alpha^{\;\;\beta}(\tau ').\end{equation}
This is a natural choice, since with this kernel a constant (gauge) $A_\mu$ results in a constant $\hat{\mathcal{A}}_\alpha$ in accordance with \eqref{eq:6}---that is, the vector potential is then constant for all observers---while it follows from \eqref{eq:10} that the electromagnetic field vanishes for all observers. A similar choice for the kernel in \eqref{eq:10} based on \eqref{eq:5}, namely,
\begin{equation}\label{eq:12} \kappa_{\alpha \beta}^{\;\;\;\; \gamma \delta}=-\frac{1}{2} (\phi_\alpha^{\;\;\gamma }\delta_\beta^{\;\;\delta} +\phi_\beta^{\;\;\delta} \delta_\alpha^{\;\;\gamma} -\phi_\beta^{\;\;\gamma}\delta_\alpha^{\;\;\delta} - \phi_\alpha^{\;\;\delta}\delta_\beta^{\;\;\gamma}),\end{equation}
cannot be the correct kernel, since that would imply that a constant electromagnetic field is measured to be constant by all (accelerated) observers in direct contradiction to experience. For instance, Kennard's experiment \cite{11}-\cite{14} is consistent with the interpretation that a nonuniformly rotating observer in a constant magnetic field measures a variable electric field \cite{15}. In this experiment, a coaxial cylindrical capacitor is inserted into a region of constant magnetic field $B_0$. The direction of the magnetic field is parallel to the axis of the capacitor. In the static situation, no potential difference is measured between the inner cylinder of radius $\rho_a$ and the outer cylinder of radius $\rho_b$. However, when the capacitor is set into rotation and a maximum rotation rate of $\Omega_{\max}$ is achieved, the potential difference (emf) in the rotating frame between the plates is found to be nonzero and in qualitative agreement with $\Omega_{\max} B_0(\rho_b^{\;\; 2}-\rho_a^{\;\; 2})/2$, which is based on the hypothesis of locality. This result is consistent with the fact that observers at rest in the rotating frame experience the presence of a radial electric field (in cylindrical coordinates). According to the hypothesis of locality, the radial electric field should have a magnitude of $\gamma vB_0$ between the cylinders, where $v=\Omega\rho$ and $\Omega$ starts from zero and reaches $\Omega_{\max}$. Here $v\ll1$; hence, $v^2$ effects can be neglected. Thus an accelerated observer in a constant magnetic field can in principle measure a variable electric field; hence, \eqref{eq:12} cannot be the field kernel in \eqref{eq:10}.

The determination of the field kernel is considerably simplified if we assume that
\begin{equation}\label{eq:13} \hat{K}_{\alpha \beta}^{\;\;\;\; \gamma \delta} (\tau ,\tau ')=\hat{k}_{\alpha \beta}^{\;\;\;\; \gamma \delta} (\tau ')\end{equation}
in accordance with \eqref{eq:4}. To generate all such kernels that are antisymmetric in their first and second pairs of indices from the Minkowski metric tensor $\eta_{\alpha \beta}$, the Levi-Civita tensor $\epsilon _{\alpha \beta \gamma \delta} $ (with $\epsilon _{0123}=1$) and the acceleration tensor $\phi_{\alpha \beta}(\tau )$ is still a rather daunting task. Examples include: (i) $W\epsilon_{\alpha \beta}^{\;\;\;\; \gamma \delta}$, where $W$ is a function of scalars formed from $\phi_{\alpha \beta}$, its dual and their time derivatives, (ii) $W\phi_{\alpha \beta}\phi^{\gamma \delta}$ and similar terms involving $\phi_{\alpha \beta}$, its dual and their time derivatives, etc. To simplify matters still much further, we forgo terms nonlinear in the acceleration tensor and concentrate instead on linear superpositions of tensor \eqref{eq:12} and its duals. The left dual of \eqref{eq:12} is given by
\begin{equation}\label{eq:14} ^\ast \kappa_{\alpha \beta}^{\;\;\;\; \gamma \delta} =\frac{1}{2}\epsilon_{\alpha \beta}^{\;\;\;\; \rho \sigma} \kappa_{\rho \sigma}^{\;\;\;\; \gamma \delta}.\end{equation}
It then follows from \eqref{eq:12} that
\begin{equation}\label{eq:15}^\ast\kappa_{\alpha \beta }^{\;\;\;\;\gamma \delta}=\frac{1}{2}(\epsilon_{\alpha\beta}^{\;\;\;\;\rho\gamma}\phi_\rho^{\;\;\delta}-\epsilon_{\alpha \beta}^{\;\;\;\;\rho\delta}\phi_\rho^{\;\;\gamma}).\end{equation}
The right dual of \eqref{eq:12}, namely,
\begin{equation}\label{eq:16}\frac{1}{2}\kappa_{\alpha \beta}^{\;\;\;\;\rho\sigma} \epsilon_{\rho \sigma}^{\;\;\;\;\gamma \delta}=-\frac{1}{2} (\phi_\alpha^{\;\;\rho}\epsilon_{\rho\beta}^{\;\;\;\;\gamma\delta}-\phi_\beta^{\;\;\rho}\epsilon_{\rho \alpha}^{\;\;\;\;\gamma\delta})\end{equation}
turns out to be equal to the left dual \eqref{eq:14}, since $\phi_{\alpha \beta}=-\phi_{\beta \alpha}$. The equality of right and left duals can be simply proved on the basis of a general identity given on page 255 of \cite{9}; indeed, our equality follows immediately from (D.1.67) of \cite{9}, since $\phi_\alpha^{\;\;\alpha}=0$. One can show that the mixed duals of \eqref{eq:12}, such as
\begin{equation}\label{eq:17} \frac{1}{2} \kappa_{\alpha \rho \sigma\beta}\epsilon^{\rho \sigma \gamma \delta},\end{equation}
all vanish \cite{15}. We therefore assume that a natural choice for kernel \eqref{eq:13} that is linear in the acceleration is given by \cite{15}
\begin{equation}\label{eq:18} \hat{k}_{\alpha \beta}^{\;\;\;\;\gamma \delta} (\tau )=p\kappa_{\alpha \beta}^{\;\;\;\;\gamma \delta} (\tau )+q\;{^\ast\kappa_{\alpha \beta}^{\;\;\;\;\gamma \delta}} (\tau ),\end{equation}
where $p$ and $q$ are constant real numbers that should be determined from observation. It follows from our discussion of Kennard's experiment that $p\neq 1$, $q\neq 0$ or both; moreover, it seems natural to assume that $p> 0$. However, the ultimate justification for the form of the kernel should be concordance with observation. With our simple choice \eqref{eq:18}, equation~\eqref{eq:10} may be written as
\begin{equation}\label{eq:19} \hat{\mathcal{F}}_{\alpha \beta}(\tau )=\hat{F}_{\alpha \beta} (\tau )+\int^\tau_{\tau_0} \kappa_{\alpha \beta}^{\;\;\;\;\gamma \delta}(\tau ') [p\hat{F}_{\gamma \delta} (\tau ')+q\hat{F}^\ast_{\gamma \delta} (\tau ')]d\tau ',\end{equation}
so that in this type of ``constitutive'' law, the nonlocal part involves a mixture of the field and its dual. This has consequences involving violations of parity and time-reversal invariance associated with a nonzero value of $q$ that are discussed in more detail in the following section. For the ``constitutive'' relation \eqref{eq:19}, it therefore appears natural to suppose that $|q|\ll1$ (cf. chapter D.1 of \cite{9}).

It is interesting to note here some properties of the tensor $\kappa_{\alpha \beta}^{\;\;\;\;\gamma \delta}$ such as $\kappa_{\alpha \beta \gamma\delta}=-\kappa_{\gamma \delta \alpha \beta}$, $\kappa^\gamma_{\;\;\alpha \gamma \beta}=-\phi_{\alpha \beta}$, $\kappa_{\alpha \beta}^{\;\;\;\;\alpha \beta}=0$ and
\begin{equation}\label{eq:20} \frac{1}{2}\kappa_{\gamma \delta}^{\;\;\;\;\rho \sigma}\kappa_{\rho \sigma}^{\;\;\;\;\gamma \delta}=-\phi_{\alpha \beta}\phi^{\alpha \beta},\quad \kappa_{\alpha \beta}^{\;\;\;\;\gamma \delta}=-\frac{1}{2} \;{^\ast \kappa_{\alpha \beta}^{\;\;\;\;\rho \sigma}} \epsilon_{\rho\sigma}^{\;\;\;\;\gamma \delta}.\end{equation}
In a similar way, one can show that $^\ast\kappa_{\alpha\beta\gamma\delta}=-^\ast\kappa_{\gamma\delta \alpha \beta}$, $^\ast \kappa^\gamma_{\;\;\alpha \gamma \beta}=-\phi^\ast_{\alpha \beta}$, etc. A general discussion of the invariants of such constitutive tensors is given in \cite{9}. To explore the implications of \eqref{eq:18} and \eqref{eq:19} for accelerated observers, we now turn to the evaluation of \eqref{eq:19} in four cases of special interest.

\section{Implications of nonlocal electrodynamics}\label{sec:7}

In the following four subsections, we explore some of the consequences of our nonlocal formulation of electrodynamics for accelerated systems. The first case involves observers rotating uniformly about the direction of propagation of an incident electromagnetic wave, which we take to be the $z$ direction, while the second case involves observers rotating nonuniformly about the direction of a constant magnetic field. In the third case, observers accelerate uniformly along the $z$ direction and measure an incident plane electromagnetic wave. The fourth case involves the measurement of a constant electromagnetic field by observers that undergo nonuniform linear acceleration.

\subsection{Uniform rotation}\label{ssec:7.1}

Consider observers that rotate uniformly with frequency $\Omega_0>0$ in planes parallel to the $(x,y)$ plane with $x=\rho\cos \varphi$, $y=\rho\sin \varphi$ and $z=z_0$, where $\rho\geq 0$ and $\varphi =\Omega_0 t=\gamma\Omega_0\tau$ for $t\geq 0$. Here $\gamma$ is the Lorentz factor corresponding to $v=\rho\Omega_0$ and we choose proper time such that $\tau=0$ at $t=0$. More specifically, imagine the class of observers that move uniformly along straight lines parallel to the $y$ axis with $x=\rho\geq 0$, $y=\rho\Omega_0t$ and $z=z_0$ for $-\infty <t<0$, but at $t=0$ are forced to move on circles about the $z$ axis. The natural orthonormal tetrad frame of such an observer for $t\geq 0$ is given by
\begin{align}\lambda^\mu_{\;\; (0)}&=\gamma (1,-v\sin \varphi ,v\cos \varphi ,0),\label{eq:21}\\
\lambda^\mu_{\;\;(1)} &=(0,\cos \varphi ,\sin \varphi ,0),\label{eq:22}\\
\lambda^\mu_{\;\;(2)} &=\gamma (v,-\sin \varphi ,\cos \varphi ,0),\label{eq:23}\\
\lambda^\mu_{\;\;(3)} &= (0,0,0,1)\label{eq:24}\end{align}
with respect to the $(t,x,y,z)$ system. In the natural cylindrical coordinate system adapted to this case, the spatial frame of the observer consists of unit vectors in the radial, tangential and $z$ directions. Moreover, the acceleration tensor can be decomposed into $\mathbf{g}=-v\gamma^2\Omega_0(1,0,0)$ and $\boldsymbol{\Omega}=\gamma^2\Omega_0(0,0,1)$ with respect to the local spatial frame \eqref{eq:22}-\eqref{eq:24}.

The observers receive a plane monochromatic electromagnetic wave of frequency $\omega$ that propagates along the $z$ direction. We use the decomposition $F_{\mu\nu}\mapsto (\mathbf{E},\mathbf{B})$ and represent the field via a column 6-vector
\begin{equation}\label{eq:25} F=\begin{bmatrix} \mathbf{E}\\ \mathbf{B}\end{bmatrix}.\end{equation}
The nonlocal theory is linear; therefore, we employ complex fields and adopt the convention that only their real parts have physical significance. Thus the incident field can be expressed as
\begin{equation}\label{eq:26} F_\pm (t,\mathbf{x})=i\omega A_\pm \begin{bmatrix}\mathbf{e}_\pm \\ \mathbf{b}_\pm \end{bmatrix} e^{-i\omega (t-z)},\end{equation}
where $A_\pm $ is a constant amplitude, $\mathbf{e}_\pm =(\hat{\mathbf{x}} \pm i\hat{\mathbf{y}})/\sqrt{2}$, $\mathbf{b}_\pm =\mp i\mathbf{e}_\pm$ and the upper (lower) sign represents positive (negative) helicity radiation. The unit circular polarization vectors $\mathbf{e}_\pm$ are such that $\mathbf{e}_\pm \cdot \mathbf{e}^\ast_\pm =1$.

Consider the measurement of the field \eqref{eq:26} by the hypothetical locally comoving inertial observers along the worldline of a rotating observer. Equation~\eqref{eq:8} may be expressed in the present notation as $\hat{F}_\pm =\Lambda F_\pm$, where $\hat{F}_{\mu\nu}\mapsto (\hat{\mathbf{E}},\hat{\mathbf{B}})$ and $\Lambda$ is a $6\times 6$ matrix given by
\begin{equation}\label{eq:27} \Lambda =\begin{bmatrix} \Lambda_1 & \Lambda_2\\ -\Lambda_2 & \Lambda_1 \end{bmatrix}\end{equation}
with
\begin{equation}\label{eq:28} \Lambda_1 =\begin{bmatrix} \gamma \cos \varphi & \gamma \sin \varphi &0\\
-\sin \varphi & \cos \varphi & 0\\ 0 & 0 &\gamma \end{bmatrix} ,\quad \Lambda_2 =v\gamma \begin{bmatrix} 0 & 0 & 1\\ 0 & 0 & 0 \\ -\cos \varphi & -\sin \varphi & 0 \end{bmatrix}.\end{equation}
Thus $\hat{F}_\pm$ can be written as 
\begin{equation}\label{eq:29} \hat{F}_\pm (\tau )=i\gamma \omega A_\pm \begin{bmatrix} \hat{\mathbf{e}}_\pm\\\hat{\mathbf{b}}_\pm \end{bmatrix} e^{-i\hat{\omega}\tau +i\omega z_0},\end{equation}
where $\hat{\mathbf{b}}_\pm =\mp i\hat{\mathbf{e}}_\pm$ and
\begin{equation}\label{eq:30} \hat{\mathbf{e}} _\pm = \frac{1}{\sqrt{2}} \begin{bmatrix} 1 \\ \pm i\gamma^{-1} \\ \pm iv\end{bmatrix}\end{equation}
are unit vectors with $\hat{\mathbf{e}}_\pm \cdot \hat{\mathbf{e}}^\ast _\pm =1$. The frequency measured in accordance with the hypothesis of locality is $\hat{\omega} =\gamma (\omega \mp \Omega_0)$ as expected.

To compute $\hat{\mathcal{F}}_\pm (\tau )$, $\hat{\mathcal{F}}_\pm \mapsto (\hat{\boldsymbol{\mathcal{E}}},\hat{\boldsymbol{\mathcal{B}}})$, corresponding to \eqref{eq:10} and \eqref{eq:19}, the components of the kernel \eqref{eq:18} have to be translated into elements of a $6\times 6$ matrix $\hat{k}$,
\begin{equation}\label{eq:31} \hat{k}=p\kappa +q\;^\ast\kappa .\end{equation}
This transformation is straightforward once factors of $2$ due to repeated indices and changes of sign due to $F_{0i} =-E_i$ are properly taken into account. To calculate the kernel, it is useful to recognize that $\Lambda$ in \eqref{eq:27} consists of $\Lambda_1$ and $\Lambda_2$ in a special arrangement; we henceforth denote this decomposition as $\Lambda \mapsto [\Lambda_1 ;\Lambda_2]$ for the sake of simplicity. It turns out that $\Lambda^{-1}$ has the same general form; in fact, $\Lambda^{-1}\mapsto [\Lambda_1^T;\Lambda_2^T]$, where $\Lambda^T_1$ is the transpose of $\Lambda_1$, etc. The result is
\begin{equation}\label{eq:32} \kappa =\begin{bmatrix} \kappa_1 & -\kappa_2\\ \kappa_2 &\kappa_1 \end{bmatrix} ,\quad {^\ast\kappa} =\begin{bmatrix} -\kappa_2 & -\kappa _1\\ \kappa_1 & -\kappa_2 \end{bmatrix} ,\end{equation}
where $\kappa_1 =\boldsymbol{\Omega} \cdot\mathbf{I}$, $\kappa_2=\mathbf{g}\cdot \mathbf{I}$ and $I_i$, $(I_i)_{jk}=-\epsilon _{ijk}$, is a $3\times3$ matrix proportional to the operator of infinitesimal rotations about the $x^i$ axis. In this way \eqref{eq:19} can be transformed into a matrix equation with the consequence that \cite{15}
\begin{equation}\label{eq:33} \hat{\mathcal{F}}_\pm (\tau )=\hat{F}_\pm (\tau ) \left[ 1+\frac{(\pm p+iq)\Omega_0 }{\omega \mp \Omega_0} (1-e^{i\hat{\omega}\tau })\right].\end{equation}
Three aspects of this result should be noted. The field measured by the accelerated observer can become a constant for any incident frequency $\omega$ such that $\omega \mp \Omega_0=-(\pm p+iq)\Omega_0$, which is impossible for $q\neq 0$. Next, for the resonance frequency $\omega =\Omega_0$ in the positive-helicity case, we find that
\begin{equation}\label{eq:34} \hat{\mathcal{F}}_+ (\tau )=\hat{F}_+ [1-i(p+iq)\gamma \Omega _0\tau ],\end{equation}
where $\hat{F}_+$ is now constant. Thus the measured field increases linearly with time and except in this special case, the measured frequency of the radiation is $\hat{\omega}=\gamma (\omega \mp \Omega_0)$ as before. Finally, it follows from \eqref{eq:33} that the ratio of the measured amplitude of positive-helicity radiation to that of negative-helicity radiation is $(A_+/A_-)\hat{\rho}$, where
\begin{equation}\label{eq:35} \hat{\rho} =\frac{\omega^2-\Omega_0^2+\Omega_0 (\omega +\Omega_0)(p+iq)}{\omega^2-\Omega_0^2-\Omega_0(\omega -\Omega_0)(p-iq)}.\end{equation}
One can show that for $\omega >\Omega_0$ and $p\geq 0$, $|\hat{\rho}|>1$. Thus for the same amplitude of incident helicities, the observer measures a higher (lower) amplitude due to nonlocality when its rotation is in the same (opposite) sense as the helicity of the incident radiation.

These consequences of acceleration-induced nonlocality for spin-rotation coupling in electrodynamics should be tested experimentally. To this end, the behavior of rotating measuring devices must be known beforehand. That is, disentangling the effect under consideration from the response of the measuring devices under rotation could be rather complicated. Such issues of principle have been treated in an interesting recent paper on the emission of radiation by a rotating atomic system \cite{16}. To proceed, we adopt a different approach based on Bohr's correspondence principle and consider electrons in the correspondence regime to be qualitatively the same as classical accelerated observers. That is, rather than directly confronting the nonlocal theory of rotating systems with observation, we study the behavior of rotating electrons in quantum theory to see which classical theory is closer to quantum mechanics in the correspondence limit \cite{17}.

In connection with the possibility that by a mere rotation of frequency $\omega$, an observer could stand completely still with respect to an incident positive helicity wave of frequency $\omega$ in accordance with the hypothesis of locality, we consider an electron in a circular ``orbit'' about a uniform magnetic field. The natural frequency for this motion is the cyclotron frequency $\Omega_c$. The transition of the electron from a given stationary state to the next one as a consequence of absorption of a photon of frequency $\Omega_c$ and definite helicity that is incident along the direction of the uniform magnetic field is studied. It follows from a detailed investigation \cite{17} that resonance occurs only for a photon of positive helicity and that in the correspondence regime $P_+\propto t^2$, while $P_- =0$, where $P_+(P_-)$ is the probability of transition for an incident positive (negative) helicity photon. This result is in qualitative agreement with \eqref{eq:34}.

The relative strength of the field amplitude measured by the rotating observer for $\omega >\Omega_0$ can  be estimated by considering the photoionization of the hydrogen atom when the electron is in a circular state with respect to the incident radiation. In this case a detailed investigation \cite{17} reveals that $\sigma_+>\sigma_-$, where $\sigma_+(\sigma_-)$ is the photoionization cross section when the electron rotates about the direction of the incident photon in the same (opposite) sense as the photon helicity. This agrees qualitatively with the fact that $|\hat{\rho}|>1$ according to \eqref{eq:35}.

\subsection{Nonuniform rotation}\label{ssec:7.2}

Imagine the same class of rotating observers as above except that for $\tau =t\geq 0$, $d\varphi /d\tau =\gamma \Omega_0 (\tau )$. Thus for $t<0$, an observer moves with constant speed $v_0=\rho\Omega_0(0)$, but for $t>0$, it has variable circular speed $v=\rho\Omega_0 (\tau )$. The orthonormal tetrad \eqref{eq:21}-\eqref{eq:24} is valid as before; however, the acceleration tensor is now characterized by
\begin{equation}\label{eq:36} \mathbf{g} (\tau )=\left(-\gamma^2 v\Omega_0,\gamma^2\frac{dv}{d\tau },0\right)\end{equation}
and $\boldsymbol{\Omega} (\tau )=\gamma^2\Omega_0 (\tau ) (0,0,1)$ with respect to the spatial tetrad frame. A detailed calculation reveals that the kernel is given by \eqref{eq:32} with $\kappa_1 =\boldsymbol{\Omega}(\tau )\cdot \mathbf{I}$ and $\kappa_2 =\mathbf{g}(\tau )\cdot \mathbf{I}$.

Suppose that a uniform static magnetic field of magnitude $B$ exists along the $z$ direction in the background frame $\mathcal{S}$. The electromagnetic field measured by the nonuniformly rotating observers is then given by
\begin{align}\label{eq:37} \hat{\mathcal{E}}_1 &= \gamma v B-p\mathcal{D} (\gamma v)B, \quad &\hat{\mathcal{E}}_2& =0, \quad &\hat{\mathcal{E}}_3&=q\mathcal{D} (\gamma )B,\\
\label{eq:38}\hat{\mathcal{B}}_1 &=-q \mathcal{D} (\gamma v)B,\quad  &\hat{\mathcal{B}}_2&=0,\quad  &\hat{\mathcal{B}}_3 &=\gamma B-p\mathcal{D} (\gamma )B,\end{align}
where. for the sake of simplicity, we have introduced the difference operator
\begin{equation}\label{eq:39} \mathcal{D} (f)=f(\tau )-f(0).\end{equation}
In the derivation of \eqref{eq:37}-\eqref{eq:38}, we have used the relations
\begin{equation}\label{eq:40}\frac{d\gamma }{d\tau }=\gamma^3v\frac{dv}{d\tau },\quad \frac{d}{d\tau } (\gamma v)=\gamma^3 \frac{dv}{d\tau }.\end{equation}

Equations~\eqref{eq:37} and \eqref{eq:38} contain the implications of nonlocality for Kennard-type experiments. Indeed, instead of the radial electric field $\gamma vB$ given by the hypothesis of locality, we find $\hat{\mathcal{E}}_1$ in \eqref{eq:37}; moreover, the measured electric field has an axial component as well given by $\hat{\mathcal{E}}_3$ in \eqref{eq:37}. The terms proportional to $q$ in \eqref{eq:37}-\eqref{eq:38} violate parity and time-reversal invariance.

\subsection{Uniform linear acceleration}\label{ssec:7.3}

Consider a linearly accelerated observer such that for $-\infty <t<0$, the position of the observer is described by $x=x_0$, $y=y_0$ and $z=z_0+v_0t$, but at $t=0$, the observer experiences a force that gives it an acceleration $g(\tau )>0$ along the positive $z$ direction. We assume that $\tau =0$ at $t=0$. The natural orthonormal tetrad of the observer is given by
\begin{equation}\label{eq:41} \lambda^\mu _{\;\;(0)}= (C,0,0,S),\quad \lambda^\mu _{\;\;(1)}= (0,1,0,0),\quad
 \lambda^\mu _{\;\;(2)}= (0,0,1,0),\quad \lambda^\mu _{\;\;(3)}=(S,0,0,C),\end{equation}
where $C=\cosh \theta$, $S=\sinh \theta$ and
\begin{equation}\label{eq:43} \theta =\theta_0 +u(\tau )\int^\tau_0g(\tau ')d\tau '.\end{equation}
Here $\theta_0$ is related to $v_0$ by $\tanh \theta_0 =v_0$ and $u(\tau )$ is the unit step function such that $u(\tau )=1$ for $\tau >0$ and $u(\tau )=0$ for $\tau <0$. The orthonormal frame of the observer is Fermi-Walker transported along its trajectory and the only nonzero components of the acceleration tensor are given by $\phi_{03}=-\phi_{30}=g$.

We assume that $g(\tau )$ is a positive constant $g_0$ until $\tau _f$ when the acceleration is turned off. That is,
\begin{equation}\label{eq:44} g(\tau )=g_0 [u(\tau)-u(\tau -\tau _f)],\end{equation}
so that the observer is uniformly accelerated for $0<\tau <\tau _f$. In this interval, we are interested in the electromagnetic measurements of the observer involving the incident plane wave given in \eqref{eq:26}. In this case, $F(\tau)$ may be written as
\begin{equation}\label{eq:45} F(\tau )=i\omega A_\pm \Theta \begin{bmatrix} \mathbf{e}_\pm\\\mathbf{b}_\pm \end{bmatrix},\end{equation} where $\Theta$ is given by \cite{18}
\begin{equation}\label{eq:46} \Theta (\theta)=e^{i\omega z_0} \exp \left[ i\frac{\omega}{g_0} (e^{-\theta}-e^{-\theta_0})\right]\end{equation}
and $\theta =\theta_0 +g_0\tau$. It turns out that the corresponding matrix $\Lambda$ has the same general form as \eqref{eq:27}; in fact, $\Lambda \mapsto [\Lambda_1 ;\Lambda_2]$, where $\Lambda_1=\diag (C,C,1)$ and $\Lambda_2 =SI_3$. Thus
\begin{equation}\label{eq:47} \hat{F}=e^{-\theta}F\end{equation}
for the locally comoving inertial observers along the worldline \cite{18}. The kernel \eqref{eq:31} can be simply computed and the result is that $\kappa$ and $^\ast\kappa$ are of the form \eqref{eq:32} with $\kappa_1=0$ and $\kappa_2=\mathbf{g}\cdot \mathbf{I}=g_0I_3$. Hence the field measured by the accelerated observer is
\begin{equation}\label{eq:48} \hat{\mathcal{F}} (\tau )=i\omega A_\pm \chi (\tau )\begin{bmatrix} \mathbf{e}_\pm\\ \mathbf{b}_\pm \end{bmatrix},\end{equation}
where $\chi$ can be expressed as
\begin{equation}\label{eq:49}\chi=e^{-\theta}\Theta +i\frac{g_0}{\omega} (p\pm iq)(\Theta -\Theta_0).\end{equation}
Here $\Theta$ is given by \eqref{eq:46} and $\Theta_0=\Theta (\theta_0)$. The term proportional to $q$ in the response function $\chi$ indicates the presence of helicity-acceleration coupling. Such a term would lead to acceleration-induced violations of parity and time-reversal invariance in electrodynamics \cite{19}.

\subsection{Nonuniform linear acceleration}\label{ssec:7.4}

We imagine the same linearly accelerated observer as above except that $g_0$ is replaced by $g_0(\tau )$ in \eqref{eq:44}. Suppose that a uniform static electromagnetic field \eqref{eq:25} exists in the background inertial frame $\mathcal{S}$. We are interested in the electromagnetic measurements of the accelerated observer for $\tau \in (0,\tau _f)$. The locally comoving inertial observers measure $\hat{F}=\Lambda F$, where $\Lambda$ has the same general form as before. We find, as expected, that
\begin{align}\label{eq:50} \hat{E}_1& =\gamma (E_1-vB_2), \quad &\hat{B}_1&=\gamma (B_1+vE_2),\\
\label{eq:51} \hat{E}_2 &= \gamma (E_2+vB_1), &\hat{B}_2&=\gamma (B_2-vE_1),\\
\label{eq:52} \hat{E}_3& =E_3, &\hat{B}_3&=B_3,\end{align}
where $\gamma=C$ and $\gamma v=S$.

The kernel in this case is given by $\kappa_1=0$ and $\kappa_2 =g_0(\tau )I_3$. It then follows from a straightforward calculation that the accelerated observer measures
\begin{align}\label{eq:53} \hat{\mathcal{E}}_1 &= \hat{E}_1-\mathcal{D}(p\hat{E}_1-q\hat{B}_1), \quad &\hat{\mathcal{E}}_2 &=\hat{E}_2 -\mathcal{D}(p\hat{E}_2 -q\hat{B}_2),\quad &\hat{\mathcal{E}}_3&=E_3,\\
\label{eq:55} \hat{\mathcal{B}}_1&=\hat{B}_1 -\mathcal{D}(p\hat{B}_1 +q\hat{E}_1),\quad & \hat{\mathcal{B}}_2 &=\hat{B}_2 -\mathcal{D}(p\hat{B}_2+q\hat{E}_2),\quad & \hat{\mathcal{B}}_3&=B_3.
\end{align}
Here the difference operator, defined in \eqref{eq:39}, has been employed. Thus the field components parallel to the direction of motion are not affected, while the components perpendicular to the direction of motion contain temporal difference terms that are proportional to $p$ and $q$. The $q$ terms violate parity and time-reversal invariance.

\section{Nonlocal Dirac equation}\label{sec:8}

According to the fundamental inertial observers at rest in the background global inertia frame $\mathcal{S}$, the Dirac equation is given by 
\begin{equation}\label{eq:57} (i\gamma^\mu\partial_\mu -m)\psi (x)=0,\end{equation}
where $\gamma^\mu$ are the constant Dirac matrices in the standard representation~\cite{20}. Here and in the rest of this paper we follow standard conventions~\cite{20}; in particular, the signature of the Minkowski metric is henceforth $-2$.

Consider an accelerated observer and the sequence of locally comoving inertial observers along its worldline. It turns out, based on a detailed analysis~\cite{21}, that according to these inertial observers the Dirac spinor is given by $\hat{\psi}(\tau)=\Lambda (\tau)\psi (\tau )$, where
\begin{equation}\label{eq:58} \Lambda (\tau)=e^{-\int_{\tau_0}^\tau\kappa_D(\tau ')d\tau '}\Lambda (\tau _0)\end{equation}
and $\kappa_D$ is an invariant matrix that can be expressed as
\begin{equation}\label{eq:59} \kappa_D (\tau )=\frac{i}{4} \phi _{\alpha \beta} (\tau )\sigma^{\alpha \beta}\end{equation}
with
\begin{equation}\label{eq:60} \sigma^{\alpha \beta}=\frac{i}{2}[\gamma^\alpha ,\gamma^\beta].\end{equation}
The Dirac spinor according to the accelerated observer $\hat{\Psi}$ is given by our nonlocal ansatz~\eqref{eq:1}. We assume that the basic requirements contained in \eqref{eq:4} and \eqref{eq:5} determine the kernel. Thus it follows from \eqref{eq:5} and \eqref{eq:58} that the Dirac kernel $\hat{k}_D$ is given by \eqref{eq:59}; that is, $\hat{k}_D=\kappa _D$ and
\begin{equation}\label{eq:61} \hat{\Psi} (\tau )=\hat{\psi} (\tau )+\int^\tau_{\tau_0} \hat{k}_D (\tau')\hat{\psi} (\tau ')d\tau '.\end{equation}

We now introduce the general notion that $\hat{\Psi}(\tau )$ may be considered to be the projection of a spinor $\Psi (\tau )$ on the local tetrad frame of the accelerated observer; that is, we write
\begin{equation}\label{eq:62} \hat{\Psi} (\tau )=\Lambda (\tau )\Psi (\tau ).\end{equation}
It follows from \eqref{eq:61} and \eqref{eq:62} that
\begin{equation}\label{eq:63} \Psi(\tau)=\psi (\tau )+\int^\tau_{\tau_0}k_D (\tau ,\tau ')\psi (\tau ')d\tau ',\end{equation}
where
\begin{equation}\label{eq:64} k_D (\tau ,\tau ')=\Lambda^{-1} (\tau )\kappa_D (\tau ')\Lambda (\tau ').\end{equation}
Furthermore, we can write, on the basis of the Volterra-Tricomi uniqueness theorems,
\begin{equation}\label{eq:65} \psi (\tau ) =\Psi (\tau )+\int^\tau_{\tau_0} r_D(\tau ,\tau ')\Psi (\tau ')d\tau ',\end{equation}
where $r_D$ is related to the resolvent kernel $\hat{r}_D$ by
\begin{equation}\label{eq:66} r_D(\tau ,\tau ')=\Lambda^{-1}(\tau )\hat{r}_D (\tau ,\tau ')\Lambda (\tau ').\end{equation}

The next step in our analysis involves the extension of \eqref{eq:63} and \eqref{eq:65} to a congruence of accelerated observers that occupy a finite spacetime domain $\Delta$ in $\mathcal{S}$. For the congruence, \eqref{eq:65} takes the form
\begin{equation}\label{eq:67} \psi (x)=\Psi (x)+\int_\Delta \mathcal{K} (x,y)\Psi (y) d^4 y,\end{equation}
where the kernel vanishes except in $\Delta$. We assume that the relationship between $\psi (x)$ and $\Psi (x)$ is unique within $\Delta$; therefore, there is a unique kernel $\mathcal{R}$ such that \eqref{eq:63} can be extended to the congruence as
\begin{equation}\label{eq:68}\Psi (x)=\psi (x)+\int_\Delta \mathcal{R}(x,y)\psi (y)d^4 y.\end{equation}
It is useful to define a nonlocal operator $\mathcal{N}$ such that
\begin{equation}\label{eq:69}\psi =\mathcal{N}\Psi,\quad \mathcal{N}\Psi (x)=\Psi (x)+\int_\Delta\mathcal{K}(x,y)\Psi (y)d^4y.\end{equation}
We note that $\mathcal{N}$ is invertible, so that $\Psi =\mathcal{N}^{-1}\psi$ is given by the right-hand side of \eqref{eq:68}. The nonlocal Dirac equation follows simply from \eqref{eq:57}, \eqref{eq:67} and \eqref{eq:69}, namely,
\begin{equation}\label{eq:70} (i\gamma^\mu \partial _\mu -m)\mathcal{N}\Psi =0.\end{equation}
It turns out that this equation in general remains nonlocal even after the acceleration of the congruence has been turned off. This circumstance is due to the persistence of the memory of past acceleration \cite{21,22,23}.

It is possible to derive \eqref{eq:70} from an appropriate action functional
\begin{equation}\label{eq:71} \mathcal{A}[\Psi ] =\int_\Delta \mathcal{L}[x,\mathcal{N}\Psi (x),\partial _\mu \mathcal{N} \Psi (x)]d^4x\end{equation}
via the principle of stationary action~\cite{24}. To this end, we recall that the Dirac equation can be obtained from the variational principle
\begin{equation}\label{eq:72} \delta\int \mathcal{L}_D d^4x=0,\end{equation}
where the local Dirac Lagrangian density is given by
\begin{equation}\label{eq:73} \mathcal{L}_D=\frac{1}{2} \bar{\psi} (i\gamma^\mu \partial_\mu -m)\psi -\frac{1}{2} [i(\partial _\mu \bar{\psi} )\gamma^\mu+m\bar{\psi}]\psi.\end{equation}
The Dirac spinor is complex; therefore, $\psi$ and its adjoint $\bar{\psi }=\psi^\dagger\gamma^0$, where $\psi^\dagger$ is its Hermitian conjugate, are varied independently in \eqref{eq:73}. The variation of the adjoint spinor results in the Dirac equation~\eqref{eq:57}, while the variation of $\psi$ results in
\begin{equation}\label{eq:74} i(\partial_\mu \bar{\psi})\gamma^\mu+m\bar{\psi}=0.\end{equation}
The two parts of the Lagrangian \eqref{eq:73} are related by a total divergence
\begin{equation}\label{eq:75} \bar{\psi }(i\gamma^\mu \partial _\mu -m)\psi =i\partial _\mu J^\mu -[i(\partial_\mu \bar{\psi })\gamma^\mu +m\bar{\psi}]\psi,\end{equation}
where $J^\mu =\bar{\psi }\gamma^\mu \psi$ is the current. This current is conserved and $\mathcal{L}_D$ vanishes once the Dirac equation is satisfied.

It has been shown \cite{24} that the desired Lagrangian in \eqref{eq:71} is obtained from the local Lagrangian \eqref{eq:73} by the nonlocal substitution of $\psi$ with $\mathcal{N}\Psi$. Thus the nonlocal Dirac Lagrangian is given by
\begin{equation}\label{eq:76} \mathcal{L}_{ND}=\frac{1}{2}\overline{\mathcal{N}\Psi} (i\gamma^\mu \partial_\mu -m)\mathcal{N}\Psi -\frac{1}{2} [i(\partial_\mu \overline{\mathcal{N}\Psi})\gamma^\mu +m\overline{\mathcal{N}\Psi}]\mathcal{N}\Psi.\end{equation}
The variation of this Lagrangian is simplified if we work, for example, with  $\delta (\mathcal{N}\Psi)=\mathcal{N}\delta \Psi$ instead of $\delta \Psi$. All such variations vanish on the boundary hypersurface $\partial \Delta$, since the kernel in \eqref{eq:69} is zero except in the finite spacetime domain $\Delta$ in which acceleration occurs.

The variational principle under consideration here involves bilinear scalar functionals of the form
\begin{equation}\label{eq:77} \langle \alpha ,\beta \rangle =\int_\Delta \bar{\alpha} \beta d^4x,\end{equation}
where $\alpha $ and $\beta$ are spinors. Moreover, for an operator $\mathcal{O}$, we define the adjoint operator $\mathcal{O}^\ast$ such that $\langle \mathcal{O} \alpha ,\beta\rangle =\langle \alpha, \mathcal{O}^\ast \beta \rangle$ with respect to the nondegenerate inner product given by \eqref{eq:77}. Thus in \eqref{eq:76}, $\overline{\mathcal{N}\Psi}$ is the adjoint spinor obtained from \eqref{eq:69} and can be expressed as
\begin{equation}\label{eq:78} \overline{\mathcal{N}\Psi} (x)=\bar{\Psi} (x)+\int_\Delta \bar{\Psi} (y)\bar{\mathcal{K}} (x,y)d^4y,\end{equation}
where
\begin{equation} \bar{\mathcal{K}}(x,y)=\gamma^0 \mathcal{K}^{\dag}(x,y)\gamma^0.\end{equation}
On the other hand, the adjoint of $\mathcal{N}$ is given by $\mathcal{N}^\ast$ such that
\begin{equation}\label{eq:80} \mathcal{N}^\ast \Psi (x)=\Psi (x)+\int_\Delta \bar{\mathcal{K}} (y,x)\Psi (y)d^4y.\end{equation}
The operator $\mathcal{N}^\ast$ is invertible because $\mathcal{N}$ is invertible \cite{25,26}. These notions are needed in order to prove that the variation of \eqref{eq:76} does indeed lead to the nonlocal Dirac equation~\eqref{eq:70}.

The nonlocal approach presented above for the Dirac equation can, in principle, be extended to other field equations. It is thus possible to derive in a consistent manner the nonlocal field equations describing interacting fields. This is illustrated in the next section, where the interaction of charged Dirac particles with the electromagnetic field is studied from the standpoint of accelerated observers.

\section{Nonlocal interaction}\label{sec:9}

Imagine a congruence of accelerated observers in the finite spacetime domain $\Delta$. We are interested in their description of the interaction of a charged Dirac particle with the electromagnetic field. Starting with the main equations of nonlocal electrodynamics \eqref{eq:9} and \eqref{eq:10} for a member of the congruence, we define the coordinate components of the fields $\mathcal{A}_\mu$ and $\mathcal{F}_{\mu\nu}$ via
\begin{equation}\label{eq:81} \hat{\mathcal{A}}_\alpha =\mathcal{A}_\mu \lambda^\mu_{\;\;(\alpha )},\quad \hat{\mathcal{F}}_{\alpha \beta}=\mathcal{F}_{\mu\nu}\lambda^\mu_{\;\;(\alpha )}\lambda^\nu_{\;\;(\beta)}.\end{equation}
Using the Volterra-Tricomi uniqueness theorems, \eqref{eq:9} and \eqref{eq:10} could be ``inverted'' by means of resolvent kernels and extended to the whole congruence thereby resulting in
\begin{align}\label{eq:82} A_\mu(x) &= \mathcal{A}_\mu (x)+\int_\Delta \mathcal{K}_\mu^{\;\;\nu}(x,y)\mathcal{A}_\nu (y)d^4 y,\\
\label{eq:83} F_{\mu\nu}(x)&=\mathcal{F}_{\mu\nu}(x)+\int_\Delta \mathcal{K}_{\mu\nu}^{\;\;\;\; \rho \sigma} (x,y)\mathcal{F}_{\rho \sigma} (y) d^4y.\end{align}
Here the kernels are assumed to vanish except in $\Delta$. Henceforth, we write \eqref{eq:82} and \eqref{eq:83} as
\begin{equation}\label{eq:84} A_\mu =n\mathcal{A}_\mu,\quad F_{\mu\nu}=N\mathcal{F}_{\mu\nu},\end{equation}
where $n$ and $N$ are invertible nonlocal operators as in \eqref{eq:69}.

The kernels $\mathcal{K}_\mu^{\;\;\nu}$ and $\mathcal{K}_{\mu\nu}^{\;\;\;\;\rho\sigma}$ in \eqref{eq:82} and \eqref{eq:83} as well as the Dirac kernel $\mathcal{K}$ in \eqref{eq:67} embody the essential nonlocality of our treatment. It is therefore necessary to explain how they are generated in nonlocal special relativity. This is done in Appendix~\ref{sec:A} for a class of uniformly accelerated observers. Further information about such kernels is contained in \cite{24}.

The Maxwell field equations for $A_\mu (x)$ and $F_{\mu\nu}(x)$ can be derived from the Lagrangian
\begin{equation}\label{eq:85}\mathcal{L}_M=\frac{1}{16\pi} F_{\mu\nu}F^{\mu\nu} -\frac{1}{8\pi} F^{\mu\nu} (\partial_\mu A_\nu -\partial_\nu A_\mu)-j_\mu A^\mu,\end{equation}
where $A_\mu$ and $F_{\mu\nu}$ are regarded as independent fields and $j_\mu$ is the current associated with charged particles. It follows from the Euler-Lagrange equations for \eqref{eq:85} that
\begin{equation}\label{eq:86} F_{\mu\nu}=\partial_\mu A_\nu -\partial _\nu A_\mu,\quad \partial_\nu F^{\mu\nu}=-4\pi j^\mu .\end{equation}
The nonlocal Maxwell equations simply follow from the substitution of \eqref{eq:84} in \eqref{eq:86}; that is,
\begin{equation}\label{eq:87} N\mathcal{F}_{\mu\nu} =\partial _\mu (n\mathcal{A}_\nu )-\partial_\nu (n\mathcal{A}_\mu ),\quad \partial_\nu (N\mathcal{F}^{\mu\nu})=-4\pi j^\mu.\end{equation}
Moreover, these equations can be derived from an action principle based on the nonlocal Maxwell Lagrangian
\begin{equation}\label{eq:88} \mathcal{L}_{NM} =\frac{1}{16\pi} (N\mathcal{F}_{\mu\nu})(N\mathcal{F}^{\mu\nu})-\frac{1}{8\pi} (N\mathcal{F}^{\mu\nu})[\partial_\mu (n\mathcal{A}_\nu)-\partial_\nu (n\mathcal{A}_\mu )]-j_\mu (n\mathcal{A}^\mu ).\end{equation}

It is important to digress here and mention that throughout this paper we have explicitly employed the standard Cartesian inertial coordinates of $\mathcal{S}$; however, it should be possible, in principle, to employ any other admissible (curvilinear) coordinate system in Minkowski spacetime. Thus the nonlocal field equations \eqref{eq:87} can be transformed to any other coordinate system based on the invariance of the forms $\mathcal{A}_\mu dx^\mu$ and $\frac{1}{2}\mathcal{F}_{\mu\nu} dx^\mu\wedge dx^\nu$. No new physical assumption is needed for this purpose; in fact, mathematical consistency is all that is required.

For the fundamental inertial observers in $\mathcal{S}$, the classical interaction of a charged Dirac particle with the electromagnetic field is expressed via the Lagrangian $\mathcal{L}=\mathcal{L}_D+\mathcal{L}_M$ with $j_\mu =\epsilon \bar{\psi} \gamma_\mu \psi$, where $\epsilon$ is the electric charge of the particle. The Euler-Lagrange equations of motion for the interacting system lead to Maxwell's equations \eqref{eq:86} together with the Dirac equation and its adjoint
\begin{align}\label{eq:89} \gamma^\mu (i\partial_\mu -\epsilon A_\mu )\psi -m\psi=0,\\
\label{eq:90} (i\partial _\mu +\epsilon A_\mu )\bar{\psi} \gamma^\mu +m\bar{\psi}=0.\end{align}
These equations imply that the current is always conserved ($\partial_\mu j^\mu =0$). Moreover, the invariance of the total action under spacetime translations leads to the conservation law
\begin{equation}\label{eq:91} \partial_\nu T^{\mu\nu}=0,\end{equation}
where $T^{\mu\nu}$ is the total canonical energy-momentum tensor of the system. That is,
\begin{equation}\label{eq:92} T^{\mu\nu}=T^{\;\;\mu\nu}_D + T_M^{\;\;\mu\nu},\end{equation}
where
\begin{equation}\label{eq:93} T_D^{\;\;\mu\nu}=\frac{i}{2} \eta^{\mu\alpha} [\bar{\psi}\gamma^\nu \partial_\alpha \psi -(\partial_\alpha \bar{\psi})\gamma^\nu \psi ]\end{equation}
and
\begin{equation}\label{eq:94} T_M^{\;\;\mu\nu}=\frac{1}{16\pi} F_{\alpha\beta} F^{\alpha \beta}\eta^{\mu\nu}-\frac{1}{4\pi} F^{\nu\alpha}\partial_\beta A_\alpha \eta^{\mu\beta}\end{equation}
are the canonical energy-momentum tensors of the free Dirac and Maxwell fields, respectively.

It follows from the preceding considerations that the corresponding acceleration-induced nonlocal interaction is described by the total Lagrangian
\begin{equation}\label{eq:95}\mathcal{L}_N=\mathcal{L}_{ND}+\mathcal{L}_{NM},\end{equation}
where $j_\mu$ in \eqref{eq:88} is given by
\begin{equation}\label{eq:96} j_\mu =\epsilon \overline{\mathcal{N}\Psi} \gamma_\mu \mathcal{N}\Psi.\end{equation}
The action principle based on \eqref{eq:95} leads to nonlocal Maxwell's equations \eqref{eq:87} with the conserved current \eqref{eq:96} as well as the nonlocal Dirac equation and its adjoint
\begin{align}\label{eq:97} \gamma^\mu (i\partial_\mu -\epsilon n\mathcal{A}_\mu )\mathcal{N}\Psi -m\mathcal{N}\Psi =0,\\
\label{eq:98} (i\partial_\mu +\epsilon n\mathcal{A}_\mu )\overline{\mathcal{N}\Psi}\gamma^\mu +m\overline{\mathcal{N}\Psi}=0.\end{align}
The nonlocal interaction terms in \eqref{eq:97} and \eqref{eq:98} are noteworthy; for inertial observers, the interaction terms arise from the replacement of the momentum operator $P_\mu =i\partial_\mu$ by $P_\mu -\epsilon A_\mu$, while for accelerated observers
\begin{equation}\label{eq:99} P_\mu\mapsto P_\mu-\epsilon [\mathcal{A}_\mu (x)+\int_\Delta \mathcal{K}_\mu^{\;\;\nu} (x,y)\mathcal{A}_\nu (y)d^4y].\end{equation}
The classical theory presented here can be extended to the quantum regime as
well, as the basic ideas of the nonlocal theory are indeed compatible with
the quantum theory.

As pointed out in \cite{24}, it is possible to define a nonlocal energy-momentum tensor $\mathcal{T}^{\mu\nu}$ that is simply obtained from \eqref{eq:92}-\eqref{eq:94} by the nonlocal substitutions \eqref{eq:69} and \eqref{eq:84}. The nonlocal equations of motion then ensure that
\begin{equation}\label{eq:100} \partial_\nu \mathcal{T}^{\mu\nu}=0.\end{equation}
It is possible to use the Belinfante-Rosenfeld procedure to obtain the
corresponding symmetric energy-momentum tensor. In fact, let $\tilde{T}^{\mu\nu}$ be the symmetric energy-momentum tensor of the interacting Dirac and Maxwell fields. That is, 
\begin{equation}\label{eq:106} \tilde{T}^{\mu\nu}=\tilde{T}_D^{\;\;\mu\nu} +\tilde{T}_M^{\;\;\mu\nu},\end{equation}
where
\begin{equation}\begin{split} \label{eq:107} \tilde{T}_D^{\;\;\mu\nu} &=\frac{1}{4} \{\bar{\psi} (\gamma ^\mu \eta ^{\nu\alpha}+\gamma^\nu \eta^{\mu\alpha})(i\partial_\alpha -\epsilon A_\alpha )\psi\\
&\quad -[(i\partial _\alpha +\epsilon A_\alpha)\bar{\psi} ](\gamma^\mu \eta^{\nu\alpha} +\gamma^\nu\eta ^{\mu\alpha})\psi\}\end{split}\end{equation}
and
\begin{equation}\label{eq:108} \tilde{T}_M^{\;\;\mu\nu}=\frac{1}{16\pi} F_{\alpha \beta} F^{\alpha \beta}\eta^{\mu\nu}-\frac{1}{4\pi}F^{\mu\alpha}F^\nu_{\;\;\alpha}.\end{equation}
Moreover, as in the canonical case, the symmetric tensor $\tilde{T}^{\mu\nu}$ is real and conserved $(\partial_\nu\tilde{T}^{\mu\nu}=0)$. The nonlocal substitutions \eqref{eq:69} and \eqref{eq:84} in $\tilde{T}^{\mu\nu}$  then lead to the symmetric nonlocal energy-momentum tensor $\tilde{\mathcal{T}}^{\mu\nu}$ such that $\partial_\nu\tilde{\mathcal{T}}^{\mu\nu}=0$. The projection of this energy-momentum tensor on the tetrad frame of an accelerated observer results in the measured components of the currents of energy and momentum. It follows from the definitions of nonlocal operators in \eqref{eq:69} and \eqref{eq:84} that these measured components are the same as those measured by the hypothetical locally comoving inertial observers. That is, the measured components of the energy-momentum tensor are in general locally defined, but for an accelerated observer these components are nonlocally related to the measured fields.

\section{Discussion}\label{sec:10}

The Minkowski spacetime of the special theory of relativity is also the arena for the nonlocal generalization of this theory. The extension of Lorentz invariance to accelerated systems is based on the hypothesis of locality. The domain of validity of this hypothesis is critically examined in this work and its application to field determination by accelerated observers is generalized to include a nonlocal contribution involving a certain average over the past worldline of the observer. Nonlocal special relativity, which reduces to the standard theory in the limit of small accelerations, has significant and interesting predictions; for instance, nonlocality forbids the existence of a fundamental scalar or pseudoscalar field. Moreover, in connection with spin-rotation coupling, it has been shown that the consequences of nonlocal electrodynamics are in better agreement with quantum mechanics in the correspondence limit than the standard theory based on the locality hypothesis. The nonlocal theory of accelerated systems appears to be in agreement with all available observational data. It would be interesting to confront the theory directly with experiment.

\section*{Acknowledgements}

I am grateful to C. Chicone, F.W. Hehl and Y.N. Obukhov for many valuable discussions.

\appendix
\section{Kernels for uniformly accelerated observers}\label{sec:A}

The purpose of this appendix is to provide examples of nonlocal kernels employed in this work for the simple case of uniformly accelerated observers.

Let us start with the primary kernel $\hat{k}(\tau )$ introduced in \eqref{eq:4}. We note that for the electromagnetic potential and the Dirac field, we have assumed that $\hat{k}$ is given by \eqref{eq:5}, while for the Faraday tensor we have instead chosen \eqref{eq:18}. It is therefore useful to define a kernel $\kappa$ such that
\begin{equation}\label{eq:A1} \frac{d\Lambda (\tau )}{d\tau }=-\kappa (\tau )\Lambda (\tau ).\end{equation}
It turns out that for \emph{uniformly} accelerated observers, $\kappa $ is constant and thus \eqref{eq:A1} has the solution
\begin{equation}\label{eq:A2} \Lambda (\tau )=e^{-\kappa (\tau -\tau _0)}\Lambda (\tau _0).\end{equation}
Thus for uniformly accelerated observers $\hat{k}$ is also constant, since in our treatment $\hat{k}=\kappa$, except in the case of Faraday tensor for which $\hat{k}=p\kappa +q\;^\ast\kappa$.

The next step is the determination of the resolvent kernel $\hat{r}(\tau ,\tau ')$. For a constant $\hat{k}$, we find that $\hat{r}$ is given in general by a convolution-type kernel (cf. Appendix~C of \cite{22})
\begin{equation}\label{eq:A3} \hat{r}(\tau ,\tau ')=-\hat{k}e^{-\hat{k}(\tau -\tau ')}.\end{equation}
We are also interested in $r(\tau ,\tau ')$,
\begin{equation}\label{eq:A4} r(\tau ,\tau ')=\Lambda^{-1} (\tau )\hat{r}(\tau ,\tau ')\Lambda (\tau ').\end{equation}
For $\hat{k}=\kappa$, equations~\eqref{eq:A2} and \eqref{eq:A3} imply that $r(\tau ,\tau ')$ is constant and can be expressed as
\begin{equation}\label{eq:A5} r=-\Lambda^{-1}(\tau _0)\kappa \Lambda (\tau _0).\end{equation}
However, if $\hat{k}\neq \kappa$, such as in \eqref{eq:31} for the Faraday tensor, equation~\eqref{eq:A4} must be worked out in detail using \eqref{eq:A2} and \eqref{eq:A3}. It is interesting to note in this connection that the matrices $\kappa$ and $^\ast\kappa$ in general commute; that is, $[\kappa ,^\ast\kappa]=0$. Moreover, $\kappa \mapsto [\kappa_1;-\kappa_2]$ and $^\ast\kappa \mapsto [-\kappa_2 ;-\kappa_1]$ result in $^\ast\kappa \kappa \mapsto [-(\kappa_1 \kappa_2 +\kappa_2\kappa_1);-(\kappa_1^2-\kappa^2_2)]$.

Finally, let us mention the extension of these results to a specific congruence of noninertial observers. Consider observers that occupy a finite open region of space $\Sigma$ and are always at rest in the background frame $\mathcal{S}$. Thus $\tau =t$ for these observers and for $-\infty <t<0$, they refer their measurements to standard inertial axes in $\mathcal{S}$. However, for $0<t<t_f$, these noninertial observers refer their measurements to axes that rotate uniformly with frequency $\Omega_0$ about the $z$ axis. The tetrad frame of these observers for $t\in (0,t_f)$ is given by \eqref{eq:21}-\eqref{eq:24} with $v=0$ and $\gamma =1$. Thus the acceleration tensor vanishes except for $t\in (0,t_f)$. For $t>t_f$, the observers refer their measurements to inertial axes that are rotated about the $z$ axis by a fixed angle $\Omega_0t_f$, as expected. We are interested in these observers for $(0,T)$, where $T\gg t_f$. Thus in this case the finite spacetime domain under consideration is $\Delta =(0,T)\times \Sigma$. The only nonzero components of the acceleration tensor are
\begin{equation}\label{eq:A6} -\phi_{12}=\phi_{21}=\Omega_0\end{equation}
and $\tau _0=0$ in this case. Thus the relevant Dirac kernel is given by \eqref{eq:A5} and \eqref{eq:59},
\begin{equation}\label{eq:A7} r_D=\frac{i}{2}\Omega_0 \sigma^{12},\end{equation}
while the relevant kernel for the electromagnetic potential is given by \eqref{eq:A5} and \eqref{eq:11}, namely,
\begin{equation}\label{eq:A8} r_\alpha^{\;\;\beta}=\phi_\alpha^{\;\;\beta}.\end{equation}
The kernel for the Faraday tensor can also be explicitly worked out in terms of $p$ and $q$ using \eqref{eq:A2} and \eqref{eq:A3}, but this will not be given here.

The extension of the Dirac kernel \eqref{eq:A7} to the congruence of noninertial observers can be expressed as
\begin{equation}\label{eq:A9} \mathcal{K}(x,x')=\mathcal{K}(t,\mathbf{x};t',\mathbf{x}'),\end{equation}
where~\cite{24}
\begin{equation}\label{eq:A10} \mathcal{K}(t,\mathbf{x};t',\mathbf{x}')=r_D\mathcal{U}_{(0,T)}(t)\mathcal{U}_{(0,t_f)} (t')\mathcal{X}_\Sigma (\mathbf{x})\delta (\mathbf{x}'-\mathbf{x}).\end{equation}
Here $\mathcal{U}_{(a,b)}$ is the unit bump function given by
\begin{equation}\label{eq:A11} \mathcal{U}_{(a,b)} (t)=u(t-a)-u(t-b).\end{equation}
Moreover, $\mathcal{X}_\Sigma$ is the characteristic function of $\Sigma$; that is, $\mathcal{X}_\Sigma (\mathbf{x})$ is unity for $\mathbf{x}$ in $\Sigma$ and zero otherwise. Similarly,
\begin{align}\label{eq:A12} \mathcal{K}_\mu^{\;\;\nu} (t,\mathbf{x};t',\mathbf{x}')&=r_\mu^{\;\;\nu} \mathcal{U}_{(0,T)}(t)\mathcal{U}_{(0,t_f)}(t')\mathcal{X}_\Sigma (\mathbf{x})\delta (\mathbf{x}'-\mathbf{x}),\\
\label{eq:A13} \mathcal{K}_{\mu\nu}^{\;\;\;\;\rho \sigma} (t,\mathbf{x};t',\mathbf{x}')&=r_{\mu\nu}^{\;\;\;\;\rho \sigma} (t,t')\mathcal{U}_{(0,T)}(t)\mathcal{U}_{(0,t_f)}(t')\mathcal{X}_\Sigma (\mathbf{x})\delta (\mathbf{x}'-\mathbf{x}).\end{align}

\end{document}